\documentclass{ws-procs975x65}

\begin{document}

\title{RADIATION FROM A $D$-DIMENSIONAL COLLISION OF SHOCK WAVES: AN INSIGHT ALLOWED BY THE $D$ PARAMETER}

\author{F. S. COELHO, C. HERDEIRO, C. REBELO and M. O. P. SAMPAIO}

\address{Departamento de F'sica da Universidade de Aveiro and I3N \\ Campus de Santiago, 3810-183 Aveiro, Portugal}

\begin{abstract}
We consider the radiation emitted in a collision of shock waves, in $D$-dimensional General Relativity (GR), and describe a  remarkably simple pattern, hinting at a more fundamental structure, unveiled by the introduction of the parameter $D$.
\end{abstract}


\bodymatter

\section{The classical trans-Planckian problem}\label{sec1}
Relativistic particle collisions are in the realm of quantum field theory. If energies are high enough such that gravity becomes relevant they should enter the realm of quantum gravity. Moreover, if a black hole forms, as expected in a trans-Planckian head-on collision, non-perturbative processes should become relevant and therefore we find ourselves with a hopeless problem in non-linear, non-perturbative, strongly time dependent quantum gravity. As first argued by 't Hooft\cite{'tHooft:1987rb} , however, well above the fundamental Planck scale this process should be well described by classical gravity (GR). The reason is that the Schwarzschild radius for the collision energy becomes much larger than the de Broglie wavelength (for the same energy) or any other interaction scale. Thus all complex quantum field theoretical interactions will be cloaked by an event horizon and therefore causally disconnected from the exterior. This horizon, in turn, will be sufficiently classical if large enough, in the sense that the quantum gravity corrections will be small on it (and therefore outside of it). In the following we shall not be concerned with how such trans-Planckian collisions could emerge as a phenomenological scenario. These have been extensively discussed in the last decade in relation with TeV gravity scenarios. We shall only assume that these collision may be well described by GR in $D$ space-time dimensions and investigate one of its fundamental properties: the inelasticity of the collision.

\section{Extracting the gravitational radiation for a head-on collision}
The gravitational field of an ultra-relativistic particle of energy $E$ is obtainable from a boost of the Schwarzschild metric. As the boost increases, the gravitational field becomes increasingly Lorentz contracted and in the limit in which the velocity goes to $c$ (keeping the energy fixed) the gravitational field (i.e. tidal forces, described by the Riemann tensor) becomes planar and has support only on a null surface; this shock wave is described by the Aichelburg-Sexl\cite{Aichelburg:1970dh} metric. Due to their flatness outside this null surface, it is possible to superimpose two oppositely moving shocks and the geometry, as an exact solution of general relativity, is completely known everywhere except in the future light cone of the collision.
Strikingly, such knowledge is enough to actually show the existence of an apparent horizon (AH) for this geometry and thus provide strong evidence that a black hole (i.e an event horizon) forms in the collision. Since the sections of the event horizon must be outside the apparent horizon, the size of the latter yields a lower bound on the size of the black hole. An energy balance argument then provides an upper bound on the amount of energy emitted in gravitational radiation in this process (i.e. the inelasticity, $\epsilon$). The computation of this bound is originally due to Penrose and was generalized to arbitrary dimension $D$ by Eardley and Giddings\cite{Eardley:2002re} , yielding
\[ \epsilon_{\rm AH \ bound}=1-\frac{1}{2}\left(\frac{D-2}{2}\frac{\Omega_{D-2}}{\Omega_{D-3}}\right)^\frac{1}{D-2} \ ,
\]
where $\Omega_n$ is the volume of the unit $n$-sphere. This bound increases monotonically with dimension approaching 50\% in the limit of infinite $D$.

Instead of computing a bound one may decide to compute the precise inelasticity by solving the Einstein equations in the future of the collision. This is a tour de force. A method, developed by D'Eath and Payne\cite{D'Eath:1992hb,D'Eath:1992hd,D'Eath:1992qu} , is to set up a perturbative approach to solve the Einstein equations. Conceptually, considering the collision in a highly boosted frame, one shock becomes much stronger than the other and the latter can be considered as a perturbation of the former. This justifies why the perturbative expansion is valid. The approach, moreover, is justified a posteriori by observing that, in linear order and in a geometric optics approximation, the radiation computed comes dominantly from a space-time region of small curvature. As a final suggestion that the method is valid, the approach yields a result (in $D = 4$) which agrees with high energy collisions in numerical relativity; the former yields $\epsilon = 16.3\%$ in second order perturbation theory\cite{D'Eath:1992qu}  (25\% in first order\cite{D'Eath:1992hb}); the latter yields $\epsilon = 14 \pm 3\%$\cite{Sperhake:2008ga} or $\epsilon=16\pm 2\%$\cite{East:2012mb} (see\cite{Cardoso:2012qm} for a review). One tantalizing observation for why second order perturbation theory should be close to the correct answer, is that in second order perturbation theory the matching between the pre-collision exact solution and post-collision perturbative solution is exact.

Extending the method of D'Eath and Payne for generic $D$ is a demanding task involving analytical and numerical methods, which we have completed, so far, to first order in perturbation theory\cite{Herdeiro:2011ck,Coelho:2012sya} (but setup the formalism to higher order\cite{Coelho:2012sy}). The inelasticity for each dimension is computed by numerical integrations of very different functions for even and odd $D$, reflecting the different properties of gravitational radiation in even dimensions (where it propagates solely on the light cone) and odd dimensions (where it propagates also inside the light-cone). At the end, remarkably, a very simple pattern emerges: the inelasticity, in first order perturbation theory, fits perfectly, i.e. with an error smaller than that of the numerical method ($< 0.1\%$), with the simple formula
\begin{equation}
\epsilon_{\rm 1st \ order}=\frac{1}{2}-\frac{1}{D} \ .
\label{for}
\end{equation}
This fit converges to $1/2$ as $D\rightarrow \infty$ just as the apparent horizon bound and it enlightens us on the $D = 4$ result: $0.25 = 1/2-1/4$. This childish observation would be insignificant solely from the four dimensional result (obtained as a numerical integration). From the pattern unveiled by considering $D$ as a parameter, however, it strongly suggests that a deeper (simpler?) understanding of the problem exists.

\section{A Balmer-like formula?}
The simplicity of the pattern exhibited by equation (\ref{for}) is mysterious, especially because of the very distinct technicalities in even and odd dimensions. Various questions arise. Do similarly simple patterns arise at higher orders? Or is this simplicity fundamentally related to linear theory? If this only occurs for the linear theory, is there a simple physical argument to compute the inelasticity in linear theory? If such simplicity holds at higher orders, what is the form of the $n^{th}$ order term? 
In such case, can the series be re-summed?

This certainly is thought stimulating and encourages further study. But the point we wish to emphasize is that the four dimensional result would never look this simple and appealing (1/4) if the $D$ dimensional computation had not been performed. A provocative thought concerning the simple pattern (\ref{for}) emerges from the following historical episode. In 1885 Johann Balmer unveiled an empirical pattern for the wavelengths of some spectral lines of the hydrogen atom. His formula (and the subsequent generalizations) formed a guiding principle in the construction of Bohr's atomic model. But the full significance of Balmer's formula was only understood with the construction of quantum mechanics. If (\ref{for}) is hinting at something deeper related to black hole formation is yet to be unveiled.

\section*{Acknowledgments}
This work was supported by grants  {\it NRHEP--295189} FP7-PEOPLE-2011-IRSES,  PTDC/FIS/116625/2010, SFRH/BD/60272/2009, SFRH/BPD/77223/2011 and SFRH/BPD/69971/2010.

\bibliographystyle{ws-procs975x65}
\bibliography{ws-pro-sample}

\end{document}